\documentclass[aps,pra,twocolumn,superscriptaddress]{revtex4-2}
\usepackage{amsfonts}
\usepackage{amsthm,amsmath,amssymb}
\usepackage{mathrsfs}
\usepackage{epsfig}
\usepackage{color}
\usepackage{graphics, graphicx}
\usepackage{bbold}
\usepackage{psfrag}
\usepackage{mathcomp}
\usepackage{subfigure}
\usepackage{verbatim}
\usepackage{color}
\usepackage[colorlinks,citecolor=blue]{hyperref}
\usepackage{csquotes}
\usepackage[utf8]{inputenc}

\begin{document}

\title{Topological phases protected by projective space-time inversion symmetry in alkaline-earth-like atoms}

\author{Xiaofan Zhou}
\affiliation{State Key Laboratory of Quantum Optics Technologies and Devices, Institute
of Laser spectroscopy, Shanxi University, Taiyuan 030006, China}
\affiliation{Collaborative Innovation Center of Extreme Optics, Shanxi University,
Taiyuan, Shanxi 030006, China}
\author{Suotang Jia}
\affiliation{State Key Laboratory of Quantum Optics Technologies and Devices, Institute
of Laser spectroscopy, Shanxi University, Taiyuan 030006, China}
\affiliation{Collaborative Innovation Center of Extreme Optics, Shanxi University,
Taiyuan, Shanxi 030006, China}
\author{Jian-Song Pan}
\email{panjsong@scu.edu.cn}
\affiliation{College of Physics, Sichuan University, Chengdu 610065, China}
\affiliation{Key Laboratory of High Energy Density Physics and Technology of Ministry of Education, Sichuan University, Chengdu 610065, China}

\begin{abstract}
An important aspect in categorizing topological phases is whether the system is spinless or spinful, given that these classes exhibit distinct symmetry algebras, leading to disparate topological classifications. By utilizing the projective presentation strategy, the topological phases of spinless (or spinful) systems can be emulated using spinful (or spinless) systems augmented with gauge fields. In this study, we propose to implement the topological phases safeguarded by the unique projective space-time inversion symmetry inherent to spinful models, using synthetic spinless alkaline-earth-like atoms. Employing the separation of orbital and nuclear-spin degrees of freedom, the model is configured as a rectangular tube penetrated by a uniform magnetic flux through each plaquette, which simulates a spinless ladder endowed with projective space-time inversion symmetry satisfying the algebraic properties of a spinful model. For interacting topological phases with inter-orbital spin-exchange interactions, which also adhere to space-time inversion symmetry, the four-fold degeneracy of edge modes is split into two pairs of edge modes with two-fold degeneracy. We map the complete phase diagram in the end and discover that these interacting topological phases ultimately evolve into distinct charge-density-wave phases via spontaneous symmetry breaking.
\end{abstract}

\maketitle


\section{Introduction}
Topological phases in free systems can generally be classified based on their internal and spatial group symmetries~\cite{thouless1982quantized,Kane2005,Bernevig2006,bernevig2006quantum,moore2007topological,konig2007quantum,
hsieh2008topological,hasan2010colloquium,qi2011topological,hasan2011three, horava2005stability, zhao2013topological, kruthoff2017topological}. The fundamental distinction in topological classification hinges on whether the systems being examined are spinful or spinless. These two spin categories exhibit distinct topological phases, originating from their differing symmetry algebra~\cite{Chiu2016}. {A prime example is the space-time inversion symmetry, denoted as $PT$, where $P$ and $T$ represent the space- and time-inversion transformation operators, respectively~\cite{PT2008,PT2016, PPT2021, PT_operators}.} In spinful systems, $(PT)^2=-1$, resulting in a Kramers double degeneracy throughout the Brillouin zone (BZ). In contrast, spinless systems have $(PT)^2=1$, ensuring a real band structure. Thus, the spin class inherently determines the topological phases that a physical system can manifest.

Recent theoretical advancements have shown that the traditional limitations in topological classification can be surpassed in the presence of gauge symmetry~\cite{PPT2021}. Specifically, the space inversion symmetry P can be projectively represented as $\mathcal{P} = GP$~\cite{projective2016,projective2020}, where $G$ is a gauge transformation. By carefully selecting $G$, the projective space-time inversion symmetry can be tailored to satisfy $(\mathcal{P}T)^2=1$ in spinful systems or $(\mathcal{P}T)^2=-1$ in spinless systems. This enables the realization of topological phases that were previously exclusive to either spinless or spinful systems in the opposite category. Notably, spinful topological phases have now been observed in inherently spinless acoustic crystals~\cite{PPT2023}. These findings open up new avenues for exploring novel topological phases beyond the conventional topological classification~\cite{Bernevig2020}.

So far, the aforementioned studies about the projective space-time inversion symmetry-protected topological phases have focused on single-particle physics, overlooking the crucial role of many-body interactions. Interactions are essential in topological phases, as they can, for example, introduce zeros in the single-particle Green's function and disrupt the bulk-boundary correspondence between topological invariants and edge states. Conversely, while additional symmetries can suppress perturbations that share those symmetries, it is believed that topological phases protected by more symmetries are more robust against perturbations due to their increased structural complexity. Therefore, investigating interacting topological phases protected by projective space-time inversion symmetry holds significant theoretical importance. Ultracold atoms in optical lattices naturally interact with each other, and their interaction strength can be precisely tuned using Feshbach resonance~\cite{manybody1,manybody2,manybody3}.
This makes ultracold atoms in optical lattices an outstanding platform for studying interacting topological phases in condensed matter physics.

In this study, we propose to realize spinful topological phases using a synthetic spinless optical lattice clock,
with synthetic dimensions implemented by nuclear and orbital degrees of
freedom~\cite{AE1,ofr1,ofr2,ofr3,ye2016old,ye2016,fallani2016,gyuboong}. 
Particularly, with the help of interactions within the clock-state manifolds controlled via the orbital Feshbach resonance~\cite{ren1,ofrexp1,ofrexp2}, we can study the impact of interaction on the topological phases protected by projective space-time inversion symmetry.
These interactions lead to the splitting of the four-fold degenerate ground-state entanglement spectra into two-fold degeneracy.
When further increasing the interaction strengthes, the interacting topological phases are driven into different order phases with spontaneous symmetry breaking. We numerically map the phase diagram and propose that the interaction-induced topological phase transitions can be explored by measuring the local occupation of clock states at the edges. Our research pioneers the exploration of topological phases protected by projective space-time inversion symmetry in the realm of ultracold atoms.

The rest of this paper is structured as follows. In section II, we elaborate the quantum simulation framework under consideration. Section III provides the definitions of the quantities employed to characterize the topological phases and phase transitions. The details of the interacting topological phases and interaction-induced phase transitions are presented in section IV. Finally, a brief summary and some discussions are given in section V.

\begin{figure}[t]
\centering
\includegraphics[width = 8.5cm]{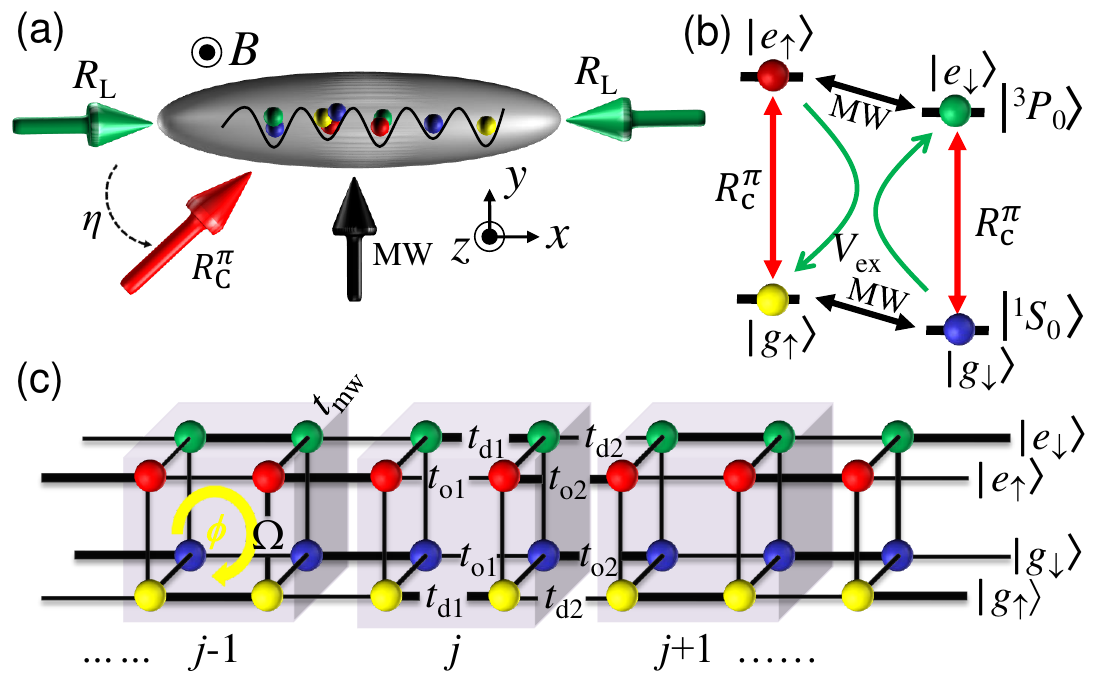} \vskip 0.0cm
\caption{
Illustration of the proposed model. (a) Ultracold alkaline-earth-like atoms are loaded onto one-dimensional (1D) spin-dependent optical superlattices created by a series of interfering laser beams denoted by $R_L$.
A narrow $\protect\pi$-polarized clock laser, $R_C^{\pi}$, induces a single-photon transition between the clock-state manifolds within the same hyperfine spin states.
By adjusting the angle $\protect\eta $ between the clock laser and the laser generating the optical lattice, the photon recoil momentum can be tuned as $k_{\mathrm{C}}=2 \pi /\lambda_{\mathrm{C}}\cos \eta$ and thus the phases carried by the coupling terms are tunable.
Microwaves (MW) are employed to couple the two hyperfine spin states.
(b) Energy level diagram and relevant coupling.
The two hyperfine spin states are coupled by microwaves.
States within the $| ^1S_0 \rangle$ and $| ^3P_0 \rangle$ manifolds are coherently coupled by the spin-conserving clock laser ($R_C^{\pi}$, indicated by the red curves).
Additionally, interactions within the clock-state manifolds can couple different spin states across different orbitals (depicted by the green curves).
(c) Configuration for the tight-binding model of the 1D optical clock. It consists of a square tube with four legs (namely, $%
| g\uparrow\rangle$, $|e\uparrow \rangle $ and $| g\downarrow
\rangle$, and $|e\downarrow \rangle $), where each on-page plaquette is penetrated by the same synthetic magnetic flux $%
\protect\phi =\protect\pi \protect\lambda _{\mathrm{L}}/\protect\lambda _{%
\mathrm{C}}\cos \protect\eta$. Gray cubes represent the unit cells of the configuration, while colored spheres represent the sites.}
\label{experimental}
\end{figure}

\section{Quantum simulation scheme}
\label{Model and Hamiltonian}
We propose to investigate the switching between interacting spinless and spinful topological phases using the following model:
\begin{eqnarray} \label{H1}
\hat{H} &\!\!\!\!=\!\!\!\!& -\!\! \sum_{j,\alpha \sigma}^{g\uparrow, e\downarrow}( t_{d1}\hat{c}_{2j-1\alpha \sigma}^{\dag} \hat{c}_{2j\alpha \sigma} + t_{d2} \hat{c}_{2j\alpha \sigma}^{\dag} \hat{c}_{2j+1\alpha \sigma} +\!\mathrm{H.c.})  \notag \\
&&- \sum_{j,\alpha \sigma}^{g\downarrow, e\uparrow} ( t_{o1} \hat{c}_{2j-1\alpha \sigma}^{\dag} \hat{c}_{2j\alpha \sigma}
+ t_{o2} \hat{c}_{2j\alpha \sigma}^{\dag} \hat{c}_{2j+1\alpha \sigma} +\!\mathrm{H.c.} ) \notag \\
&&+\Omega\sum_{j,\sigma }(e^{i\phi j}\hat{c}_{jg\sigma }^{\dag }\hat{c}_{je\sigma
}\!+\!\mathrm{H.c.}) + t_{\rm mw}\sum_{j,\alpha }(\hat{c}_{j\alpha\uparrow }^{\dag }\hat{c}_{j\alpha\downarrow}\!+\!\mathrm{H.c.})  \notag \\
&&+U\sum_{j}(\hat{n}_{jg\uparrow }\hat{n}_{je\downarrow }+\hat{n}%
_{jg\downarrow }\hat{n}_{je\uparrow })+U_{0}\sum_{j\sigma }\hat{n}_{jg\sigma
}\hat{n}_{je\sigma }  \notag \\
&&+V_{\mathrm{ex}}\sum_{j}(\hat{c}_{jg\uparrow }^{\dag }\hat{c}%
_{je\downarrow }^{\dag }\hat{c}_{je\uparrow }\hat{c}_{jg\downarrow }+\mathrm{%
H.c.}),
\end{eqnarray}
where $\hat{c}_{j\alpha \sigma }$ ($\hat{c}_{j\alpha \sigma }^{\dag }$) denotes the annihilation (creation) operator for an atom on the jth site of the $\alpha=e,g$ orbital with spin $\sigma=\uparrow, \downarrow$, and $\hat{n}_{j\alpha \sigma }=\hat{c}
_{j\alpha \sigma }^{\dag }\hat{c}_{j\alpha \sigma}$. This Hamiltonian is formulated in the context of alkaline-earth-like atoms on an optical clock lattice. Note that the term \enquote{spinless} does not imply the absence of atomic spin degrees of freedom; rather, it indicates that the Hamiltonian adheres to a space-time inversion symmetry algebra of spinless systems, i.e., $(PT)^2=1$, where the two hyperfine spin states are treated as two spatial degrees of freedom, constituting a synthetic spatial dimension. Consequently, the Hamiltonian describes a one-dimensional spatial tube [Fig.~\ref{experimental}(c)]. Due to staggered longitudinal hopping, each unit cell comprises 8 sites. The $\Omega$ hopping terms include a site-dependent phase, equivalent to a constant flux piercing the on-page plaques. The interaction terms are associated with orbital Feshbach resonances.

An optical clock lattice can be created by loading atoms into a series of one-dimensional (1D) spin-dependent optical superlattice potentials. They can be achieved by interfering laser beams ($R_L$) at specific 'tune-out' wavelengths [see Fig.~\ref{experimental}(a)], ensuring that atoms in the state $\alpha,\sigma$ are exclusively confined by the lattice potential $V^{\mathrm{L}}_{\alpha,\sigma}$ and not influenced by other potentials~\cite{Arora2011,zhangNature}. The spin- and orbital-dependence is accomplished by using lasers with different frequencies and circular polarizations to couple the hyperfine states in distinct orbitals to higher energy levels. The superlattice is formed by the superposition of a background lattice and an additional lattice with twice the lattice constant.

The spin-conserved $\Omega$ term, is realized using an ultranarrow $\pi$-polarized clock laser ($R_C^{\pi}$) with a wavelength $\lambda _{\mathrm{C}}$ [see Fig.~\ref{experimental}(a)]. This laser drives a single-photon transition between the clock states $|g \rangle$ and $| e\rangle$, which share the same nuclear spins $\sigma$ ($| \!\! \uparrow \rangle$ or $| \!\! \downarrow \rangle$) [see Fig.~\ref{experimental}(b)]. The angle $\eta$ between the wave vector of the clock laser and the 1D optical lattice [see Fig.~\ref{experimental}(a)] results in a momentum transfer of $k_{\mathrm{C}}=2\pi/\lambda _{\mathrm{C}} \cos \eta $. By adjusting $\eta$, we can tune the site-dependent phase of the $\Omega$ terms. To couple the two spin states $| \!\! \uparrow \rangle$ and $| \!\! \downarrow \rangle$ [see Fig.~\ref{experimental}(b)], we propose using microwaves (MW)~\cite{zhangNature} or two-photon Raman processes~\cite{mancini2015observation}. When the 1D optical lattice is sufficiently deep and the Rabi frequency of the clock laser is not excessively large~\cite{highband1,highband2,highband4}, the single-band approximation becomes valid, allowing us to formulate the tight-binding model as described above.

Under this setup, the longitudinal spin-conserving hopping coefficients are given by $t_{\alpha\sigma,jj+1}=\left\vert \int
dx{w}^{(j)}\left[ -\frac{\nabla ^{2}}{2m}+V^{\mathrm{L}}_{\alpha, \sigma}\left(
x\right) \right] w^{(j+1)}\right\vert $, with $w^{(j)}$ being the lowest-band Wannier
function at the $j$th site of the background lattice potential. Here we have assumed that the background lattice potentials and thus the Wannier functions are the same for different orbitals and hyperfine spins. The double-well potentials are designed to obtain $t_{g\uparrow,2l 2l+1}=t_{e\downarrow,2l 2l+1}=t_{o1}$, $t_{e\uparrow,2l 2l+1}=t_{g\downarrow,2l 2l+1}=t_{d1}$, $t_{g\uparrow,(2l+1)(2l+1)+1}=t_{e\downarrow,(2l+1)(2l+1)+1}=t_{o2}$, and $t_{e\uparrow,(2l+1)(2l+1)+1}=t_{g\downarrow,(2l+1)(2l+1)+1}=t_{d2}$.
The spin-conserving coupling coefficient $\Omega e^{i\phi j}$ {$=\Omega _{%
\text{R}}\int dxw^{(j)}e^{ik_{\mathrm{C}}x}w^{(j)}$ }, in which $%
\Omega _{\text{R}}$ is the Rabi frequency of the clock laser.
Here $\phi =\frac{1}{2}k_{\mathrm{C}}\lambda _{\mathrm{L}}=\pi
\lambda _{\mathrm{L}} /\lambda _{\mathrm{C}}\cos \eta$ is the synthetic
flux penetrating each on-page plaquette, where $\lambda_{L}$ is the wavelength of the lasers generating the background lattice. For the interaction part, $U=\frac{1}{2}\left(
g_{+}+g_{-}\right) \int dxw^{(j)}w^{(j)}w^{(j)}w^{(j)}$, and $U_{0}=${$%
    g_{-}\int dxw^{(j)}w^{(j)}w^{(j)}w^{(j)}$} measures the inter-orbital
density-density interaction strengths with the same and different nuclear
spins, respectively. $V_{\mathrm{ex}}=\frac{1}{2}\left(
g_{-}-g_{+}\right) \int dxw^{(j)}w^{(j)}w^{(j)}w^{(j)}$ is the inter-orbital
spin-exchange interaction strength.

The Hamiltonian (\ref{H1}) offers the advantage of allowing independent tuning of all its parameters. Specifically, $t_{\alpha\sigma,jj+1}$ can be adjusted by modifying the depth of the optical lattice potential, while $\Omega $ and $\phi $ can be controlled via the Rabi frequency and the angle of the clock laser, respectively. It has been reported that $\Omega$ terms can be realized in experiment~\cite{fallani2016}.  Additionally, the parameters $\{V_{\mathrm{ex}}$, $U$, $U_{0}\}$ can be fine-tuned using orbital Feshbach resonance~\cite{ren1,ofrexp1,ofrexp2} or confinement-induced resonance~\cite{CIR,ren2}, where the relevant scattering lengths can be tuned from $-\infty$ to $\infty$. Subsequently, we set $U_0=V_{\mathrm{ex}}+U$, as determined by the scattering parameters of $^{173}$Yb atoms~\cite{CIR,ren2}.

The single-particle part of Hamiltonian (\ref{H1}) is formulated based on the one-dimensional single-particle models presented in Refs.~\cite{PPT2021,PPT2023}, thereby inheriting the same symmetrical framework. We fix the flux $\phi=\pi$ to ensure that each plaquette perpendicular to the $x$ direction experiences a $\pi$ flux. The coupling strengths, given by $\Omega e^{i\phi j}$, alternate between positive and negative values for neighboring links, specifying a $\mathbb Z_2$ gauge field where $\mathbb Z_2= \{ \pm1 \}$. A gauge transformation $G$ applied to this $\mathbb Z_2$ gauge field, through $\hat{c}_{j\alpha\downarrow}\rightarrow-\hat{c}_{j\alpha\downarrow}$, maintains the flux configuration intact.

The space inversion symmetry operation (P) is executed by interchanging the spins $\uparrow$ and $\downarrow$, the orbitals $g$ and $e$, and the sites $j$ and $-j$. It's worth emphasizing that, in this framework, the two spin degrees of freedom are regarded as spatial degrees of freedom. As a result, the space-time inversion symmetry adheres to the relation $(PT)^2=1$, which is characteristic of spinless space-time inversion symmetric models. Although the flux configuration is unchanged for the spatial reversion transformation, the gauge connection and thus the Hamiltonian are changed. In fact, due to the presence of the $\mathbb Z_2$  gauge symmetry, the Hamiltonian submits to the projective space inversion symmetry $\mathcal{P}=GP$. However, the projective space-time inversion symmetry fulfills $(\mathcal{P}T)^2=-1$, thereby determining the spinful algebra and topological classifications, although we start with a synthetic spinless model.

The 1D topologically gapped phase belongs to the DIII class of spinful models. We set the parameters as follows: $t_{d1}=t_{o2}=t-\Delta_t$, $t_{d2}=t_{o1}=t+\Delta_t$, $\Omega=0.1$ and $t_{\rm mw}=0.1$, with $t=0.2$. According to the experiment on the lattice potential of $^{173}$Yb atoms with $759$nm lasers~\cite{fallani2016}, the average hopping coefficient $t$ can be tuned up to several tens of times the recoil energy. A double-well lattice can be created by combining a background standing wave with another standing wave that has a double wavelength~\cite{atala2014observation}. In principle, the additional double-wave-length standing wave can be realized using a laser with the same frequency but tilted at 30 degrees. The ratio between $t$ and $\Delta_{t}$ can be tuned by varying the intensity ratio between the background and additional standing waves. The lattices for different spin and orbital degrees of freedom can be specified with the polarization and frequencies of lasers. The Hamiltonian (\ref{H1}) features four Kramers double-degenerate bands and two Majorana Kramers pairs of topological boundary modes, which are localized at one end of the chain and guaranteed by the projective space-time inversion symmetry with $(\mathcal{P}T)^2=-1$.
The winding number of the chain is constrained to even integers, denoted as 2Z~\cite{PPT2021}. The band gap closes at $\Delta_t=0$, marking a topological phase transition. To study the topological insulator properties of interacting fermions, we focus on the case where $\Delta_t=0.1$.
At half filling, where $\rho=N/4L=0.5$, the lower four bands and two zero modes are occupied, resulting in the ground state exhibiting topological insulating behavior with edge states. To quantitatively explore the topological properties of the model, we employ the state-of-the-art DMRG numerical method~\cite{dmrg1,dmrg2}. We consider a lattice length up to $L=56$ at half filling ($\rho=0.5$). For our calculations, we retain 400 truncated states per DMRG block and perform 30 sweeps, achieving a maximum truncation error of approximately $\sim$ $10^{-10}$.

\section{Quantities for characterizing interacting topological phases}
\label{order parameters}
The topological states and phase transitions in strongly correlated systems generally can be characterized by the entanglement spectrum, entanglement entropy, and excited energy gap. The entanglement spectrum is defined as follows~\cite{Li2008}:
\begin{equation}
\xi_{i} = -\ln(\rho_{i}),
\end{equation}
where $\rho_{i}$ represents the eigenvalues of the reduced density matrix $\hat{%
\rho}_{l}=\mathrm{Tr}_{L-l}|\psi \rangle \langle \psi |$, with $|\psi \rangle$ being the ground state. Here, $l$ denotes the length of the left block in a specific bipartition. A system is considered topologically nontrivial if its entanglement spectrum exhibits degeneracy, as this degeneracy is indicative of the presence of edge excitations~\cite{Li2008,Zhao2015,Yoshida2014,Turner2011,Pollmann2010,
Fidkowski2010,Flammia2009}.

The quantum criticality of interaction-driven topological phase transitions can be characterized by the von Neumann entropy~\cite{Flammia2009, Hastings2010, Daley2012, Abanin2012, Jiang2012, Islam2015}:
\begin{equation}
S_{\mathrm{vN}} = -\mathrm{Tr}_{l}[\hat{\rho}_{l} \log \hat{\rho}_{l}],
\end{equation}
where $l=L/2$. A divergence in the von Neumann entropy at the critical point signifies a continuous transition. The central charge of the conformal field theory describing the critical behavior can be extracted from the asymptotic behavior of the entanglement entropy, which reflects the universality class of the phase transition. For a critical system with open boundary conditions, the von Neumann entropy of a subchain of length $l$ scales as:
\begin{equation}
S_{\mathrm{vN}}(l) = \frac{c}{6} \ln \left[ \sin \left( \frac{\pi l}{L} \right) \right] + \text{const},
\end{equation}
where the slope at large distances gives the central charge $c$ of the conformal field theory~\cite{centralcharge1, centralcharge2, centralcharge3, centralcharge4}.

For the TI phase under open boundary conditions, a notable characteristic is the emergence of localized edge modes. These topological edge modes manifest within the bulk energy gap and can be identified from the chemical potential spectrum corresponding to different occupations. The chemical potential spectrum is defined as:
\begin{equation}
\mu(N) = E_0(N + 1) - E_0(N),
\end{equation}
representing the energy needed to add an atom to a system comprising $N$ atoms. Here, $E_0(N)$ denotes the ground-state energy for a system with $N$ atoms under open boundary conditions. The density distribution of the edge modes can be computed using:
\begin{equation}
\Delta \hat{n}_{j\alpha \sigma} = \langle \hat{n}_{j\alpha \sigma}(N+1) \rangle - \langle \hat{n}_{j\alpha \sigma}(N) \rangle,
\end{equation}
where $\langle\hat{n}_{j\alpha \sigma}(N)\rangle$ represents the density distribution for a system with $N$ atoms.

The exchange interactions can lead to various order states with spontaneous symmetry breaking, including the rung-singlet states, the charge-density wave (CDW) state, and the orbital-density wave (ODW) state, by evaluating their respective local orders~\cite{phases1,phases2,phases3}. The singlet states within the orbital (or spin) degrees of freedom are defined as $|\pm\rangle=(|g \! \uparrow;e \! \downarrow\rangle \pm |g \! \downarrow;e \! \uparrow\rangle)/\sqrt{2}$ and can be distinguished through the corresponding local order parameter:

\begin{equation}\label{RS_order}
\langle \hat{\rho}_{j\pm}\rangle = |\pm\rangle\langle\pm|.
\end{equation}


The analytical expression of topological invariant for the free case of the topological phases we consider has been given in Ref.~\cite{PPT2021}. Due to the lack of well-defined Bloch energy bands, topological invariants of interacting systems are typically not so clear to the community~\cite{hohenadler2013correlation}. To the best of our knowledge, there is no a unified framework for writing down the topological invariants for all topological classifications in presence of interaction. In this paper, although we can confirm the existence of topological phases using quantities like entanglement spectra, topological edge modes, and the variations of energy gaps, it is beyond our ability to write down the expression for topological invariant in presence of interactions, although it is also important.

\section{Interaction driven topological phase transitions}

\begin{figure}[t]
\centering
\includegraphics[width = 8.8cm]{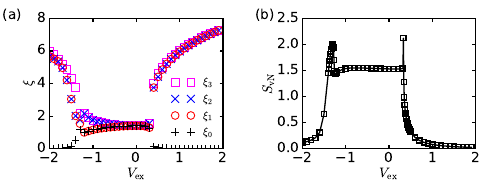} \hskip 0.0cm
\caption{Variations of entanglement spectrum and von Neumann entropy. (a) The lowest four levels in the entanglement spectrum $\protect\xi_i$ ($i=1,2,3,4$), and (b) the von Neumann entropy $S_{\mathrm{vN}}$ as functions of $V_{\rm ex}$ with $U=0.0$, $L=56$ and $\rho=0.5$ under open boundary conditions.}
\label{fig:phase transition}
\end{figure}

\begin{figure}[b]
\centering
\includegraphics[width = 8.8cm]{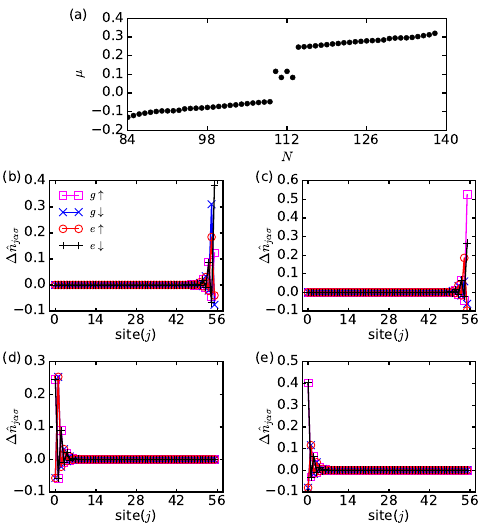} \hskip 0.0cm
\caption{Chemical potential and edge modes. (a) Chemical potential $\protect\mu(N)$ for a chain with $L=56$ lattice sites under open boundary conditions. (b)-(e) The edge-mode density distributions $\Delta \hat{n}_{j\alpha \sigma}$ of the four mid-gap modes for the TI state.
Here, $V_{\mathrm{ex}}=0.2$ and $U=0.0$.}
\label{fig:phases TI}
\end{figure}

In the absence of interactions, the system behaves as a topological insulator (TI). Here, we investigate the effect of increasing the spin-exchange interaction $V_{\mathrm{ex}}$ while keeping $U$ fixed at zero. {As shown in Fig.~\ref{fig:phase transition}(a)}, when $V_{\mathrm{ex}}=0$, the eigenstates in the entanglement spectrum exhibit a fourfold degeneracy. However, this degeneracy is partially lifted in the presence of a weak $V_{\mathrm{ex}}$. As the interaction strength $V_{\mathrm{ex}}$ is further increased, the degeneracy of $\xi_{i}$ is completely lifted beyond a critical repulsive (attractive) interaction strength of $V^{c}_{\mathrm{ex}} \sim 0.35$(-1.25), as illustrated in Fig.~\ref{fig:phase transition}(a). { This signals topological phase transitions from an interacting topological phase
to trivial phases.} Additionally, the von Neumann entropy $S_{\mathrm{vN}}$ displays sharp peaks near these two critical points, as shown in Fig.~\ref{fig:phase transition}(b).

\begin{figure}
\centering
\includegraphics[width = 8.8cm]{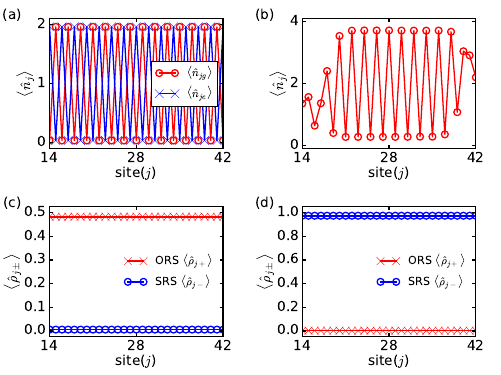} \hskip 0.0cm
\caption{Order parameters induced by interaction. (a) The density distribution $\langle\hat{n}_{j\alpha}\rangle$ in ODW phase with $V_{\text{ex}}=0.0$ and $U=1.8$.
(b) $\langle\hat{n}_{j}\rangle$ in CDW phase with $V_{\text{ex}}=-0.2$ and $U=-0.6$.
(c) The local order parameter $\langle \hat{\rho}_{j\pm}\rangle$ in ORS phase with $V_{\text{ex}}=-1.9$ and $U=1.0$.
(d) $\langle \hat{\rho}_{j\pm}\rangle$ in the SRS phase with $V_{\text{ex}}=1.5$ and $U=0.15$.
Here, $L=56$ and $\rho=0.5$.}
\label{fig:phases}
\end{figure}

For the single-particle topological insulator (TI) state, mid-gap modes corresponding to the four topological edge modes (two for each orbital) appear in the chemical-potential spectrum. These mid-gap modes signify the excitation energies needed as the occupation number of the edge modes incrementally rises from zero to four. Nonetheless, when there is a finite interaction strength, such as $V_{\mathrm{ex}}=0.2$, these mid-gap modes shift towards the bulk, as depicted in Fig.~\ref{fig:phases TI}(a). As the interaction strength increases further, the mid-gap modes ultimately merge into the bulk spectrum. The edge modes exhibit localized density distributions at the edges, as illustrated in Figs.~\ref{fig:phases TI}(b) through \ref{fig:phases TI}(e). { The edge modes shown in Fig.~\ref{fig:phases TI}(b) and (d) (Fig.~\ref{fig:phases TI}(c) and (e)) are degenerate and locate at different boundaries, respectively.}

\begin{figure}[b]
\centering
\includegraphics[width = 5.5cm]{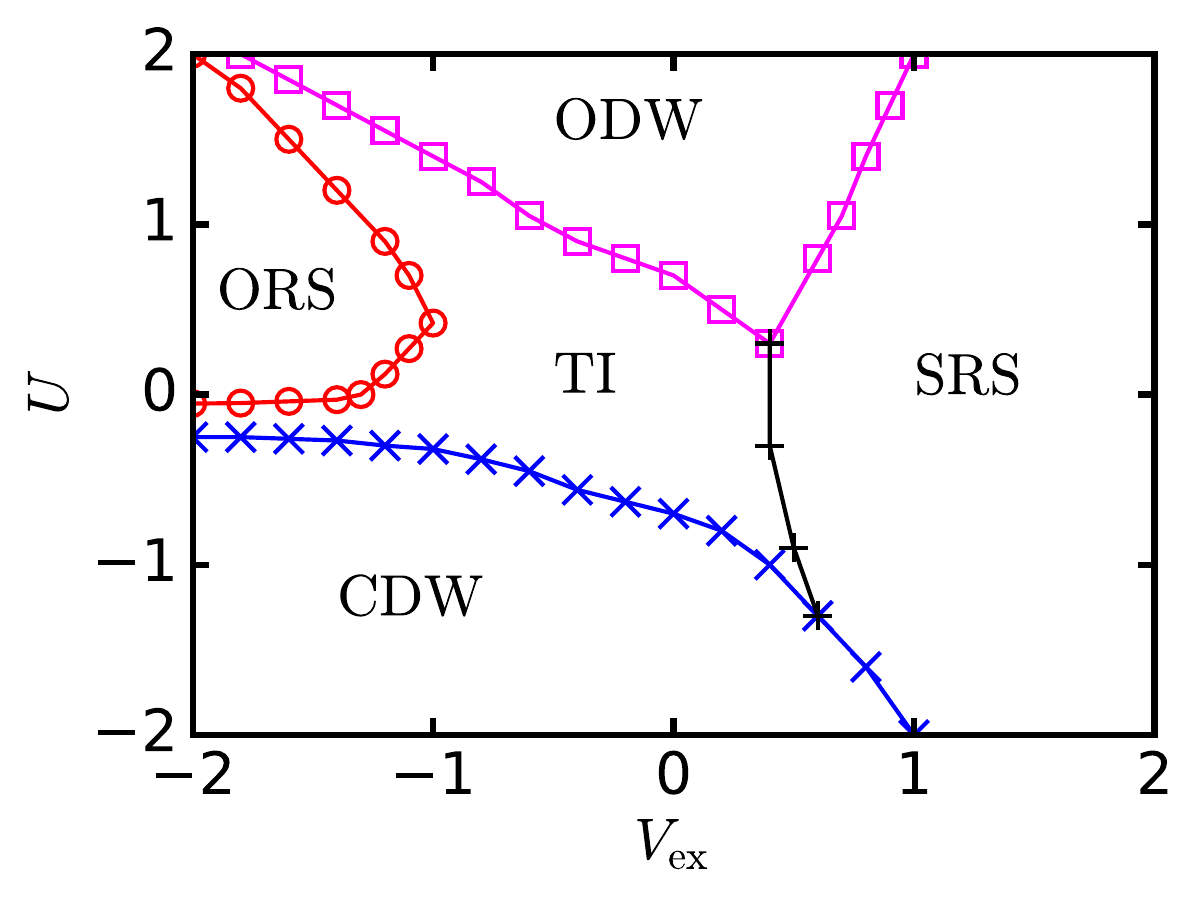} \vskip 0.0cm
\caption{The phase diagram of Hamiltonian (\ref{H1}) on $V_{\mathrm{ex}}-U$ plane with half filling $\rho=0.5$ in the thermodynamic limit, which contains TI (topological insulator), SRS (spin rung-singlet), ORS (orbital rung-singlet), ODW (orbital density wave) and CDW (charge density wave) phases.}
\label{fig:phase diagram}
\end{figure}

As the interaction strengths are further increased, multiple trivial order phases arise. Figure~\ref{fig:phases} indicates that, for the repulsive $V_{\rm ex}$ case with $U=0$, the trivial symmetric state is the spin rung-singlet (SRS) state~\cite{phases3}. This state can be described by the direct-product state $\prod_i |-\rangle_i$. When $U$ becomes non-zero, the system can transition into the orbital rung-singlet (ORS) state ($\prod_i |+\rangle_i$), the charge-density wave (CDW) state, or the orbital-density wave (ODW) state, depending on the topological phase boundaries. Both the CDW and ODW states are trivial ordered states characterized by spontaneously broken chiral symmetry, which can be verified by computing the relevant local quantities as shown in Fig.~\ref{fig:phases}.

\begin{figure}[b]
\centering
\includegraphics[width = 8.8cm]{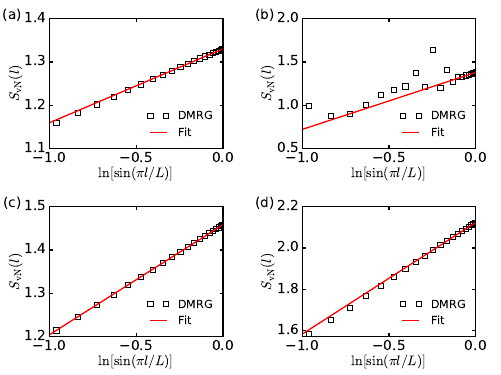} \hskip 0.0cm
\caption{The scaling of von Neumann entropy $S_{\mathrm{vN}}(l)$ as function of $\sin(\pi l/L)$ at critical points.
The red solid lines are the linear fit with: $S_{\rm vN}=c/6\ln[\sin(\pi l/L)]+$const.
The central charge $c$ is six times the slope of the linear fit.
(a) The T$\rightarrow$ODW phase boundary with $V_{\rm ex}=0.0$ and $U=0.51$, $c \sim 1.0$.
(b) The T$\rightarrow$CDW phase boundary with $V_{\rm ex}=0.0$ and $U=-0.514$, $c \sim 3.94$.
(c) The T$\rightarrow$ORS phase boundary with $V_{\rm ex}=-1.0$ and $U=0.6$, $c \sim 1.5$.
(d) The T$\rightarrow$SRS phase boundary with $V_{\rm ex}=0.335$ and $U=0.0$, $c \sim 3.2$.
Here, $L=56$ and $\rho=0.5$.}
\label{fig:critical}
\end{figure}

Utilizing the entanglement spectrum $\xi_{i}$, von Neumann entropy $S_{\mathrm{vN}}$, order parameter calculations, and finite-size scaling, we have constructed the phase diagram in the thermodynamic limit, as depicted in Fig.~\ref{fig:phase diagram}. { For weak and moderate interaction strength, the ground-state of the system behaves an interacting topological phase. The repulsive interorbital spin-exchange interaction $V_{\mathrm{ex}}$ can drive the topological phase to SRS phase, but the attractive $V_{\mathrm{ex}}$ favors ORS phase. The repulsive (attractive) interorbital density-density interaction $U$ favors the ground-state ODW (CDW) phase, respectively. These charge-density-wave-like phases with similar profiles in the phase diagram have been found previously in one of our works with the same form of interaction terms~\cite{zhou2017symmetry}. This is mainly due to that the phase boundaries between these charge-density-wave-like phases are mainly determined by the competitions between interaction terms.} The numerical results suggest that the critical points for phase transitions remain consistent across various finite lattice sizes L.

The system behaves as Luttinger liquids at the phase boundaries. As illustrated in Fig.~\ref{fig:critical}, by analyzing the divergence of the von Neumann entropy at these boundaries, we have determined that the central charges for the T-ODW, T-CDW, T-ORS, and T-SRS phase transitions are approximately 1.0, 3.94, 1.5, and 3.2, respectively. These critical points delineate several lines that serve as the phase boundaries between the quantum phases. These lines represent Luttinger liquids with varying central charges, as depicted in Fig.~\ref{fig:phase diagram}.

\section{Conclusions and discussions}
\label{Conclusions}
In this paper, we first present a detailed quantum simulation scheme for exploring spinful topological phases in optical lattices of synthetic spinless alkaline-earth-like atoms. {Our model is constructed by coupling four Su-Schrieffer-Heeger (SSH) chains in a form that submits to the projective space-time inversion symmetry, which thus can simulate topological phases in DIII class that can not be realized in spinless model, when we map the model to a spinful model}. Next, we numerically characterize the topological phase with DMRG numerical calculation. {In our scheme, contact and spin-exchange interactions naturally exist and can be tuned with Feshbach resonance, and we carefully discuss the impact of interactions.} Our findings reveal that interactions lift the four-fold degeneracy of edge modes into a two-fold degeneracy. As interaction strengthes further increase, various order phases with spontaneous symmetry breaking arise. Lastly, we present a comprehensive phase diagram that encompasses both the interacting spinful topological phase and the interaction-induced order phases.

{ We would like to emphasize the essential differences between our current work and one of our previous studies~\cite{zhou2017symmetry}. The model in Ref.~\cite{zhou2017symmetry} is essentially formed by two coupled one-dimensional spin-$1/2$ chains with spin-orbit coupling (SOC)~\cite{liu2013manipulating}. The two SOC chains are coupled through interaction terms. For the SOC chain, the time-reversal and charge-conjugate symmetries are broken and the topological phases are only protected by the chiral symmetry. It thus belongs to AIII class in the ten-fold Altland-Zirnbauer classification. In contrast, the model studied here is formed by four coupled SSH chains. The four SSH chains are coupled not only by interaction terms but also by inter-spin and inter-orbital coupling terms. The SSH model belongs to the BDI class. The four SSH chains are coupled in a form that preserves the protective space-time inversion symmetry, and its topological phases are not only protected by global symmetry for internal degrees of freedom but also by space-time symmetries. The topological phases of our model thus belong to a subset of BDI class, which is mapped to DIII class when we map the model to a spinful model~\cite{PPT2021,PPT2023}. In principle, different topological classes have different topological protection mechanisms. Therefore, both the model and the topological phases proposed here are all essentially different from those studied in our previous work Ref.~\cite{zhou2017symmetry}.

The similarity between the profiles of phase diagrams in our current work and previous work~\cite{zhou2017symmetry} is mainly due to the fact that the interaction terms are set to be the same in the two works, and the competition between interaction terms dominates the phase boundaries of the charge-density-wave-like phases. Nevertheless, the critical differences emerge upon examining parameter regimes where the two models diverge substantially. For instance, in Ref.~\cite{zhou2017symmetry}, the topological phase persists even under strong negative exchange interactions ($V_{ex}\ll-t_s$) when onsite interactions vanish ($U=0$). Conversely, in the current model, the topological phase collapses at $V_{ex}~-6t$ when $U=0$. Therefore, the model and topological phases studied here are all essentially different from those studied in Ref.~\cite{zhou2017symmetry}.
}

\section*{Acknowledgments}
X.Z. and S.J. are supported by National Key Research and Development Program of China under Grant No.
2022YFA1404201, the National Natural Science Foundation of China (NSFC) under Grant Nos.~12174233, 12004230
and 12034012, Fundamental Research Program of Shanxi Province under Grants No. 202403021221024,
the Research Project Supported by Shanxi Scholarship Council of China and Shanxi ``1331KSC''.
J.S.P. is supported by the Natural Science Foundation of Sichuan Province under Grant No. 2025ZNSFSC0058, the Science Specialty Program of Sichuan University under Grant No. 2020SCUNL210, the Fundamental Research Funds for the Central Universities under Grant No. YJ202212 and the National Key R$\&$D Program of China under Grant No. 2024YFF0508503.


\begin{thebibliography}{99}
\bibitem{PT_operators}{In the field of quantum physics, the space-time inversion symmetry is also referred to as parity-time symmetry.}
\end{thebibliography}
\end{document}